\documentclass[aps,prl,twocolumn,groupedaddress]{revtex4}
\usepackage{epsfig}

\begin{document}

\def\d{{\rm d}}
\def\eps{\varepsilon}
\def\lp{\left. }
\def\rp{\right. }
\def\lr{\left( }
\def\rr{\right) }
\def\le{\left[ }
\def\re{\right] }
\def\lg{\left\{ }
\def\rg{\right\} }
\def\lb{\left| }
\def\rb{\right| }
\def\beq{\begin{equation}}
\def\eeq{\end{equation}}
\def\bea{\begin{eqnarray}}
\def\eea{\end{eqnarray}}
\def\mr{\,\mathrm{}}

\preprint{DESY 13-182}
\preprint{MS-TP-13-34}
\title{NNLO contributions to jet photoproduction and determination of $\alpha_s$}
\author{Michael Klasen$^a$}
\email[]{michael.klasen@uni-muenster.de}
\author{Gustav Kramer$^b$}
\author{Markus Michael$^a$}
\affiliation{$^a$ Institut f\"ur Theoretische Physik, Westf\"alische
 Wilhelms-Universit\"at M\"unster, Wilhelm-Klemm-Stra\ss{}e 9, D-48149 M\"unster,
 Germany\\
 $^b$ II.\ Institut f\"ur Theoretische Physik, Universit\"at Hamburg, Luruper
 Chaussee 149, D-22761 Hamburg, Germany}
\date{\today}
\begin{abstract}
We present the first calculation of inclusive jet photoproduction
with next-to-next-to-leading order (NNLO) contributions, obtained from a
unified threshold resummation formalism. The leading coefficients
for direct photoproduction are computed analytically. Together with
the coefficients pertinent to parton-parton scattering, they are shown to
agree with those appearing in our full next-to-leading order calculations.
For hadron-hadron scattering, numerical agreement is found with a previous
calculation of jet production at the Tevatron. We show that the direct
and resolved NNLO contributions considerably improve the description of
final ZEUS data on jet photoproduction and that the error on the
determination of the strong coupling constant is significantly
reduced.
\end{abstract}
\maketitle

\vspace*{-84mm}
\noindent DESY 13-182 \\
\noindent MS-TP-13-34 \\
\vspace*{61mm}


\section{Introduction}

The HERA collider, which operated at DESY from 1992 to 2007, has
produced many important physics results, first of all perhaps the most
precise determinations to date of the quark and gluon densities in the
proton from single experiments (H1, ZEUS) \cite{Adloff:2000qk,%
Chekanov:2001qu} and their combined data sets \cite{Aaron:2009aa}. These
data, taken in deep-inelastic electron-proton scattering, are
complemented by a wealth of data from photoproduction at low virtuality
of the exchanged photon, in particular on jet production, giving access
also to the distributions of partons in the photon and to measurements
of the strong coupling constant \cite{Klasen:2002xb}.

Using the full data set of the HERA run from 2005-2007 with an integrated
luminosity of 300 pb$^{-1}$, the ZEUS collaboration have recently published
a final measurement of inclusive jet photoproduction \cite{Abramowicz:2012jz}
and used it to determine the strong coupling constant (at the mass $M_Z$ of
the $Z$-boson) to be
\beq
 \alpha_s(M_Z)=0.1206^{+0.0023}_{-0.0022} ({\rm exp.}) ^{+0.0042}_{-0.0035} ({\rm th.}),
 \label{eq:1}
\eeq
based on a comparison with our next-to-leading order (NLO) QCD calculations
\cite{Klasen:1995ab}. 
While this value (like the one obtained from deep-inelastic electron-photon
scattering \cite{Albino:2002ck}) is in agreement with the current world average
of $\alpha_s(M_Z)=0.1184\pm0.0007$ \cite{Beringer:1900zz},
it is also less precise, since the latter uses only observables that
are known to next-to-next-to-leading order (NNLO) of perturbative QCD.

In this Letter, we compute the inclusive jet photoproduction cross
section for the first time including NNLO contributions, obtained
from a unified threshold resummation formalism \cite{Kidonakis:2003tx},
and extract the first NNLO value for the strong coupling constant from
photoproduction data. Our calculations are based on our previous work
on inclusive jet \cite{Klasen:1994bj} and dijet \cite{Klasen:1995ab}
photoproduction and lead, as we will see, to a considerably improved
description of the ZEUS data and a theoretical error on $\alpha_s$ that
is significantly reduced.

\section{NNLO contributions to jet photoproduction}
\label{sec:2}

The QCD factorization theorem allows to write the differential cross section for
inclusive jet photoproduction
\bea
 \d\sigma&=&
 \sum_{a,b}
 \int\d y         \,f_{\gamma/e}(y)
 \int\d x_{\gamma}\,f_{a/\gamma}(x_{\gamma},\mu_{\gamma})\nonumber\\ &\times& \hspace*{5mm}
 \int\d x_p       \,f_{b/p}(x_p,\mu_p)
 \ \d\sigma_{ab}(\alpha_s,\mu,\mu_\gamma,\mu_p)
\eea
as a convolution of the partonic scattering cross section
$\d\sigma_{ab}$, which includes both direct ($a=\gamma$, $b=q,g$) and
resolved ($a,b=q,g$) photon contributions, with the Weizs\"acker-Williams
flux of photons in electrons $f_{\gamma/e}$ and the parton densities in the
photon and proton $f_{a/\gamma}$ ($\delta(1-x_{\gamma})$ for direct photons)
and $f_{b/p}$, respectively.

From a unified threshold resummation formalism a master formula
can be obtained that allows to compute soft and virtual corrections
to arbitrary partonic hard scattering cross sections \cite{Kidonakis:2003tx}.
At NLO it reads
\bea
 \d\sigma_{ab}&=&\d\sigma_{ab}^B{\alpha_s(\mu)\over\pi}
 \le c_3D_1(z)+c_2D_0(z)+c_1\delta(1-z)\re \nonumber\\ &+&
 {\alpha_s^{d_{\alpha_s}+1}(\mu)\over\pi}\le A^cD_0(z)+T^c_1\delta(1-z)\re,
\eea
where the second line is only present for processes with complex color
flow (here resolved processes), $d_{\alpha_s}$ denotes the power of $\alpha_s$
already present in the Born term $\d\sigma_{ab}^B$ (1 for direct
and 2 for resolved photoproduction), and
\bea
 D_l(z)&=&\le{\ln^l(1-z)\over 1-z}\re_+
\eea
with decreasing $l$ are the leading and subleading logarithms at
partonic threshold ($z\to1$) in pair-invariant-mass kinematics.
The NNLO master formula is given in the reference cited above,
as are the general and complex color flow formul\ae\ for the
coefficients $c_i$, $A^c$ and $T_1^c$.

The coefficients for the simple color flow in direct photoproduction
are given here for the first time. For the QCD Compton process
$\gamma q\to qg$, we find $c_3=C_F-N_C$,
\bea
 c_2 &=& C_F\le-\ln\lr{\mu_p^2\over s}\rr-{7\over4}+\ln\lr{-u\over s}\rr\re\nonumber\\
     &+& {N_C\over2}\le-1+\ln\lr{t\over u}\rr\re-{\beta_0\over4},
\eea
and the scale-dependent part of the coefficient
\bea
 c_1^\mu&=&-{3C_F\over4}\ln\left(\frac{\mu_p^2}{s}\right)
           +\frac{\beta_0}{4}\ln\left(\frac{\mu^2}{s}\right).
\eea
For the crossed process $\gamma g\to q\bar{q}$, we find $c_3=2(N_C-C_F)$,
\bea
 c_2 &\!=\!& -{5C_F\over2}
      +  N_C\le-\ln\lr{\mu_p^2\over s}\rr-{1\over2}+{1\over2}\ln\lr{tu\over s^2}\rr\re,\quad
\eea
and the scale-dependent part of the coefficient
\bea
 c_1^\mu &=& -{\beta_0\over4}\ln\lr{\mu_p^2\over s}\rr
             +{\beta_0\over4}\ln\lr{\mu^2\over s}\rr.
\eea
For both processes,
the scale-independent parts of $c_1$ can be found in Refs.\ \cite{Klasen:1995ab}.
These coefficients depend on the QCD color factors $C_F=4/3$ and $N_C=3$,
the one-loop $\beta$-function $\beta_0=(11N_C-2n_f)/3$ with $n_f$ quark
flavors, the Mandelstam variables $s$, $t$ and $u$, and the  renormalization and proton
factorization scales $\mu$ and $\mu_p$, but not on the photon factorization
scale $\mu_\gamma$, as the QED splitting of a photon to a quark-antiquark
pair is not enhanced by the logarithms given above. We therefore only expect the
dependencies on the former two scales to be improved at NNLO.
We have verified that at NLO the direct and resolved
coefficients agree with the virtual and soft initial-state corrections
calculated in pair-invariant-mass kinematics for two jets integrated
analytically over singular and numerically over regular regions of
phase space \cite{Klasen:1995ab}.
At NNLO, these coefficients appear exclusively in the three leading logarithmic
terms $D_3$, $D_2$ and $D_1$ of the master formula. For lower
terms, only the scale-depenent terms are known and included, but the
resulting error should be small.

\section{Comparisons with D0, ZEUS and H1 data}
\label{sec:3}

\begin{figure}
\centering
\includegraphics[width=\columnwidth]{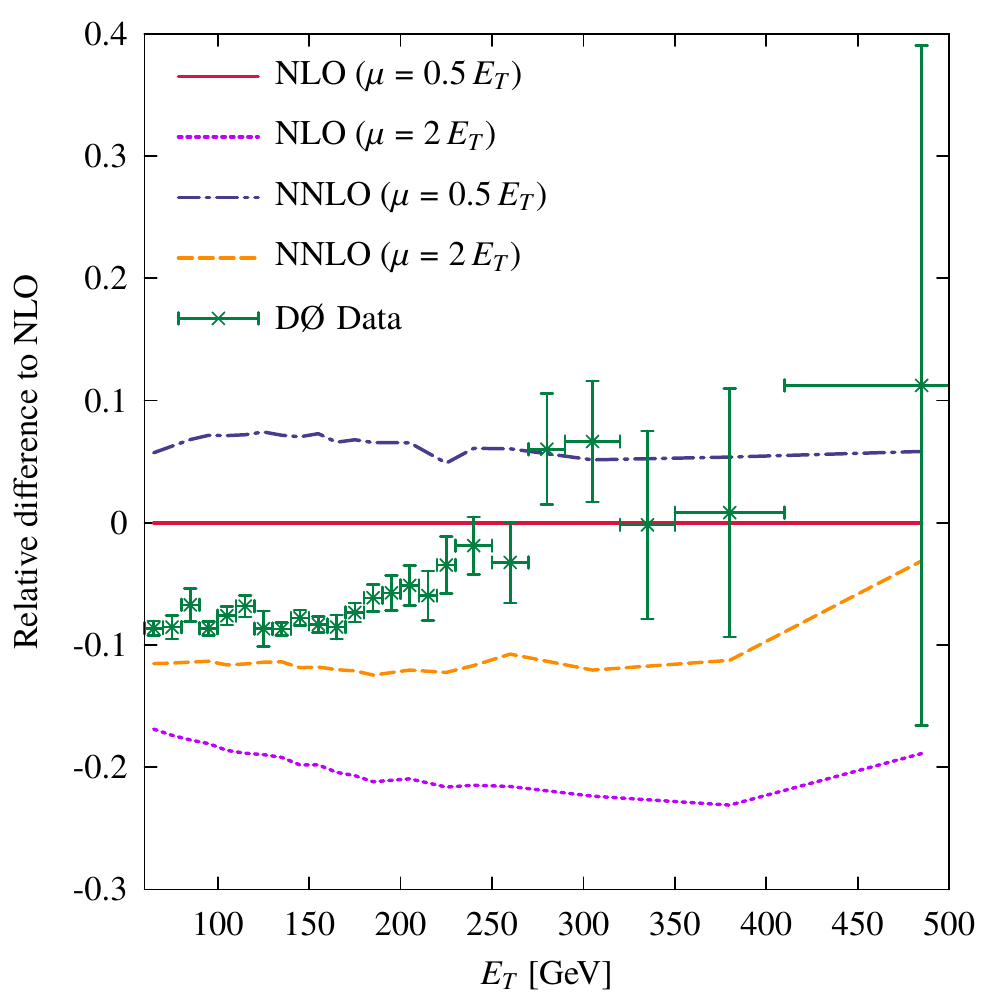}
\caption{\label{fig:1} Relative difference of D0 data, NLO and
 NNLO predictions with scales $\mu,\mu_p=[0.5;2]\times
 E_T$ to the central NLO prediction as a function of jet transverse
 energy $E_T$.}
\end{figure}

The analytical results described above have been implemented in
our program for dijet photoproduction, where a convolution over $z$ was
already performed for initial-state singularities \cite{Klasen:1995ab}.
The NNLO terms for simple color flow were implemented exclusively on the
proton side, while those for complex color flow (resolved photoproduction)
have been split evenly among the photon and the proton. At NLO, we use of
course our complete calculation and not only the logarithmically enhanced
terms described above. 
As a numerical check, we have repeated the calculation of inclusive jet
production at the Tevatron at NLO and NNLO with different scales as shown
in Fig.\ 2 of Ref.\ \cite{Kidonakis:2000gi} (note that the rapidity
range there should read $|\eta|\leq0.5$), finding full agreement (see our
Fig.\ \ref{fig:1}).
%
%

The final ZEUS measurements for inclusive jet photoproduction and the
determination of $\alpha_s$ were performed with photon virtuality $Q^2<1$
GeV$^2$, $\gamma p$ center-of-mass energy in the range $142<W_{\gamma p}<293$
GeV, and using the inclusive $k_T$-algorithm \cite{Catani:1993hr} with jet
radius $R=1$, transverse energy $E_T>17$ GeV and rapidity in the range
$-1<\eta<2.5$ \cite{Abramowicz:2012jz}. To facilitate easy comparison of
the following figures, we also employ the ZEUS-S \cite{Chekanov:2002pv}
and GRV-HO \cite{Gluck:1991ee} fits of the parton densities in the proton
and photon, where the latter has been transformed from the DIS$_{\gamma}$
to the $\overline{\rm MS}$ scheme. Similar measurements have also been
performed almost a decade earlier by the H1 collaboration
\cite{Adloff:2003nr}. Due to the lower integrated luminosity available
then (24.1 pb$^{-1}$), the data are less precise, but we have verified
that within errors they are in good agreement with our calculations at
NLO and NNLO.

\begin{figure}
\centering
\includegraphics[width=\columnwidth]{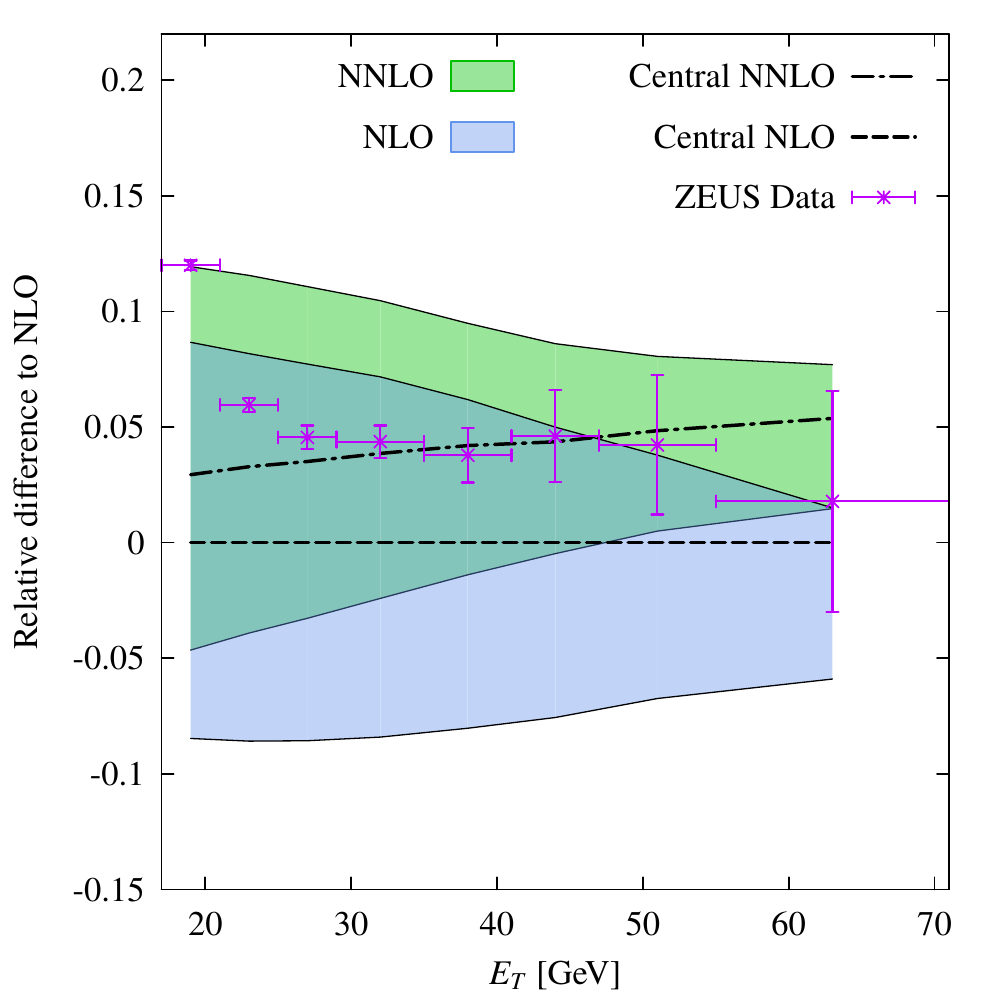}
\caption{\label{fig:2} Relative difference of ZEUS data, NLO and
 NNLO predictions with scales $\mu,\mu_\gamma,\mu_p=[0.5;2]\times
 E_T$ to the central NLO prediction as a function of jet transverse
 energy $E_T$.}
\end{figure}

Focusing now on the more precise ZEUS data, we compare in Fig.\
\ref{fig:2} the measured and various theoretical transverse energy
%
%
spectra to the NLO prediction with central scales $(\mu,\mu_{\gamma},
\mu_p=E_T)$ and after applying hadronization corrections \cite{Abramowicz:2012jz}.
The NLO uncertainty band (blue) is obtained by varying
the scales about the central scale up and down by a factor of two
and coincides with the one shown in Fig.\ 2 of Ref.\
\cite{Abramowicz:2012jz}. The NNLO corrections (including NNLO running
of $\alpha_s$) increase the
central prediction by 3-6\%, bringing it into considerably better
agreement with the experimental data. As expected from the general
behavior of threshold logarithms, the increase is larger at high
$E_T$. The scale uncertainty is also reduced at NNLO (green band),
in particular at high $E_T$, where it drops from 8 to 5\%.
Note that the data point in the largest $E_T$-bin has been omitted
from this comparison, as it suffers from large experimental (in
particular jet energy-scale) uncertainties.

\begin{figure}
\centering
\includegraphics[width=\columnwidth]{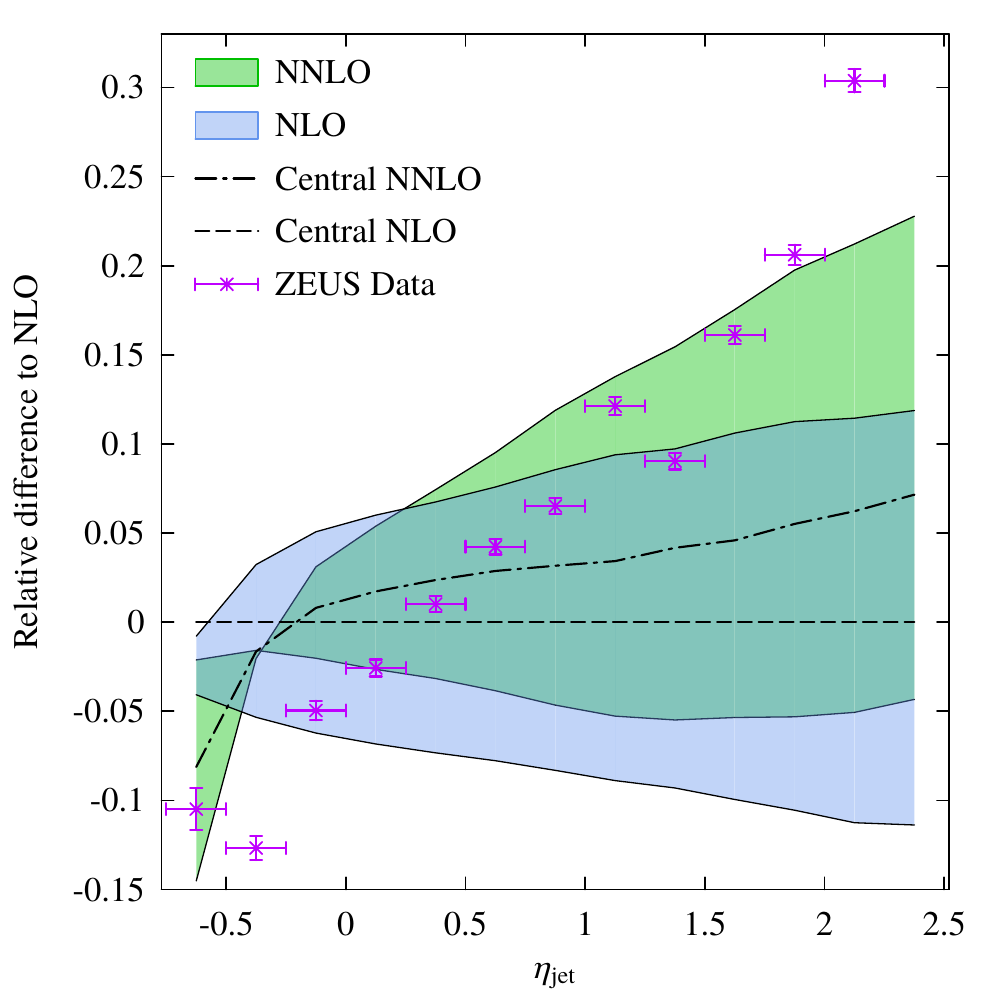}
\caption{\label{fig:3} Relative difference of ZEUS data, NLO and NNLO
 predictions with scales $\mu,\mu_\gamma,\mu_p=[0.5;2]\times E_T$ to the
 central NLO prediction as a function of jet rapidity $\eta$.}
\end{figure}

In Fig.\ \ref{fig:3} we perform a similar comparison for the
%
%
jet rapidity distribution. This distribution has been problematic
since the earliest HERA runs, as it tended to be overestimated in the
backward (photon) and underestimated in the forward (proton) direction.
These discrepancies are indeed observed at NLO in Fig.\ \ref{fig:3},
together with a large scale uncertainty, in particular in the forward
direction. They were traditionally assigned to hadronization, but
also missing higher-order corrections. This conjecture can now be
corroborated for the first time, as the NNLO contributions do bring
the theoretical predictions into better agreement with the data in
both kinematical regions. The scale uncertainty is not significantly
improved at NNLO, as the data are dominated by the low-$E_T$ region,
but within it they can now be described up to the largest rapidities.
Note that all figures in this section have been obtained using the
world average value for $\alpha_s(M_Z)=0.118$ as required for the
ZEUS-S (standard) fit \cite{Chekanov:2002pv}.

\section{Determination of $\alpha_s$}
\label{sec:4}

To determine the strong coupling constant from these comparisons, the
theoretical calculations have to be performed with a set of parton
densities in the proton obtained from global fits assuming different
values of $\alpha_s(M_Z)$. For our analysis at NNLO, we employ the
latest fits of the CTEQ-TEA collaboration (CT10), which have been obtained
with NNLO running of the coupling, evolution of the parton densities, deep-inelastic
scattering and vector-boson production matrix elements \cite{Gao:2013xoa}.
13 different CT10 NNLO sets were used, which correspond to values of
$\alpha_s(M_Z)=0.112$ to 0.124. In contrast, the ZEUS determination at
NLO in Eq.\ (\ref{eq:1}) was based
on only five different sets of ZEUS-S parton densities corresponding to
values of $\alpha_s(M_Z)=0.115$ to 0.123. Its theoretical error was
dominated by higher-order terms, estimated from
scale variations, but depended only weakly on variations of the
parton densities in the proton, the hadronization model, and the jet
algorithm.

At NLO, for a scale choice of $\mu,\mu_{\gamma},\mu_p=E_T$, and
omitting the lowest $E_T$ point, which lies clearly outside the theoretical
error band at this order (see Fig.\ \ref{fig:2}), we reproduce
$\alpha_s(M_Z)=0.121^{+0.002}_{-0.002} ({\rm exp.}) ^{+0.005}_{-0.003} ({\rm th.})$
as in Eq.\ (\ref{eq:1}) and the ZEUS analysis \cite{Abramowicz:2012jz}.
As stated there, including the lowest $E_T$ point severely worsens the
quality of the fit, and we find an increase in the minimal value of $\chi^2$/d.o.f.\
from 16/6 to 123/7.
Through this result, we also confirm that the fit of the strong coupling
constant does not
depend very much on the employed parton densities in the proton, which were
CT10 NNLO in our fit and ZEUS-S NLO in the ZEUS analysis.
The uncertainty induced by the parton densities in the
photon was systematically studied by the ZEUS collaboration and
estimated to be ${+2}/{-1}$ \%. It is expected to remain the same
at NLO and NNLO, in particular due to the lack of more
precise deep-inelastic electron-photon scattering data and a NNLO
fit to them.

Finally, at NNLO and for the same scale choice as the one given above, we obtain
\beq
 \alpha_s(M_Z)=0.120^{+0.002}_{-0.002} ({\rm exp.})^{+0.003}_{-0.003}({\rm th.}).
\eeq
The central value is now lower, since the NNLO contributions increase the
cross section for all $E_T$ bins (see Fig.\ \ref{fig:2}), and is brought
closer to the world average of 0.118. In addition, the theoretical error
is significantly reduced, which reflects the stabilization of the
cross section prediction with respect to variations of the unphysical scales.

\section{Conclusions}
\label{sec:5}

In conclusion, we have presented here a first calculation
of inclusive jet photoproduction up to NNLO of perturbative QCD.
Leading and subleading logarithmic contributions were extracted from
a unified threshold resummation formalism for photon-parton and parton-parton
scattering processes pertinent to direct and resolved photoproduction of jets
and shown to agree with those appearing in our full NLO calculations.
The NNLO contributions implemented in our NLO program were shown to correctly
reproduce results obtained in the literature for hadron-hadron scattering
at the Tevatron and to considerably improve the description of
final ZEUS data on jet photoproduction. A NNLO fit of these data with the
CT10 set of parton densities resulted in a new determination of
the strong coupling constant at the mass of the $Z$-boson
in agreement with the current world average and the ZEUS determination at NLO,
but with a significantly reduced theoretical error.

\acknowledgments

We thank C.\ Glasman, N.\ Kidonakis, and W.\ Vogelsang for useful discussions.
This work has been supported by the BMBF Theorie-Verbund ``Begleitende
theoretische Untersuchungen zu den Experimenten an den Gro\ss{}ger\"aten der
Teilchenphysik.''



\begin{thebibliography}{00}


\bibitem{Adloff:2000qk} 
  C.~Adloff {\it et al.}  [H1 Collaboration],
  Eur.\ Phys.\ J.\ C {\bf 21}, 33 (2001).

\bibitem{Chekanov:2001qu} 
  S.~Chekanov {\it et al.}  [ZEUS Collaboration],
  Eur.\ Phys.\ J.\ C {\bf 21}, 443 (2001).

\bibitem{Aaron:2009aa} 
  F.~D.~Aaron {\it et al.}  [H1 and ZEUS Collaboration],
  JHEP {\bf 1001}, 109 (2010).


\bibitem{Klasen:2002xb} 
  M.~Klasen,
  Rev.\ Mod.\ Phys.\  {\bf 74}, 1221 (2002).


\bibitem{Abramowicz:2012jz} 
  H.~Abramowicz {\it et al.}  [ZEUS Collaboration],
  Nucl.\ Phys.\ B {\bf 864}, 1 (2012).


\bibitem{Klasen:1995ab} 
  M.~Klasen and G.~Kramer,
  Z.\ Phys.\ C {\bf 72}, 107 (1996) and
%
  Z.\ Phys.\ C {\bf 76}, 67 (1997);
%
  M.~Klasen, T.~Kleinwort and G.~Kramer,
  Eur.\ Phys.\ J.\ direct C {\bf 1}, 1 (1998).


\bibitem{Albino:2002ck} 
  S.~Albino, M.~Klasen and S.~S\"oldner-Rembold,
  Phys.\ Rev.\ Lett.\  {\bf 89}, 122004 (2002).

\bibitem{Beringer:1900zz} 
  J.~Beringer {\it et al.}  [Particle Data Group Collaboration],
  Phys.\ Rev.\ D {\bf 86}, 010001 (2012).


\bibitem{Kidonakis:2003tx} 
  N.~Kidonakis,
  Int.\ J.\ Mod.\ Phys.\ A {\bf 19}, 1793 (2004);
  see also
  D.~de Florian and W.~Vogelsang,
  Phys.\ Rev.\ D {\bf 76}, 074031 (2007)
  and
  M.~C.~Kumar and S.~O.~Moch,
  arXiv:1309.5311 [hep-ph].



\bibitem{Klasen:1994bj} 
  M.~Klasen, G.~Kramer and S.~G.~Salesch,
  Z.\ Phys.\ C {\bf 68}, 113 (1995);
%
  M.~Klasen and G.~Kramer,
  Phys.\ Rev.\ D {\bf 56}, 2702 (1997).


\bibitem{Kidonakis:2000gi} 
  N.~Kidonakis and J.~F.~Owens,
  Phys.\ Rev.\ D {\bf 63}, 054019 (2001).


\bibitem{Catani:1993hr}
  S.~Catani, Y.~L.~Dokshitzer, M.~H.~Seymour and B.~R.~Webber,
  Nucl.\ Phys.\ B {\bf 406}, 187 (1993);
%
  S.~D.~Ellis and D.~E.~Soper,
  Phys.\ Rev.\ D {\bf 48}, 3160 (1993).


\bibitem{Chekanov:2002pv} 
  S.~Chekanov {\it et al.}  [ZEUS Collaboration],
  Phys.\ Rev.\ D {\bf 67}, 012007 (2003).

\bibitem{Gluck:1991ee} 
  M.~Gl\"uck, E.~Reya and A.~Vogt,
  Phys.\ Rev.\ D {\bf 45}, 3986 (1992) and
%
  Phys.\ Rev.\ D {\bf 46}, 1973 (1992).


\bibitem{Adloff:2003nr} 
  C.~Adloff {\it et al.}  [H1 Collaboration],
  Eur.\ Phys.\ J.\ C {\bf 29}, 497 (2003).


\bibitem{Gao:2013xoa} 
  J.~Gao, M.~Guzzi, J.~Huston, H.~-L.~Lai, Z.~Li, P.~Nadolsky, J.~Pumplin and D.~Stump {\it et al.},
  arXiv:1302.6246 [hep-ph].

\end{thebibliography}
\end{document}